\documentclass[preprint,floatflix,amsmath]{revtex4}

\usepackage{graphicx}

\usepackage{amssymb}

\begin{document}




\title{A non arbitrary definition of rain event: the case of stratiform rain}


\author{M.Ignaccolo$^{1}$, C. De Michele$^{2}$ \\
1) Duke University, FEL, Department of Physics, Durham, NC, United States of America
2)University of North Carolina, DESE, Chapel Hill, NC, United States of America}
\date{\today}
\begin{abstract}
A long standing issue in Hydrology is the arbitrariness of the rain ``event'' definition. In this manuscript, we show that 
1) the event definition resting on the occurrence of a minimum rainless period and the one resting a sequence 
of consecutive wet intervals are statistically equivalent. 2) In the case of stratiform rain, a non arbitrary definition of rain event is possible. 
The dynamical properties of stratiform rain indicate the range $[$1.5,4$]$h as the proper one for the choice of a minimum rainless 
period for Chilbolton, UK. 3) The intra event dynamical variability is ``described'' by an alternate sequence of quiescent and active phases. 
\end{abstract}



\section{Introduction}\label{intro}
The importance of a reliable statistical definition of rain event resides in the possibility of creating viable models 
for a great variety of hydrological applications: e.g. catchment run-off, soil erosion and canopy losses \cite{DD08}. 
Ecohydrologic studies rely on rain event modelling to study the influence of the precipitation pulse dynamics on the
vegetation architecture, plant species interactions, and  the cycling of organic matter and nutrients in arid and semi-arid 
ecosystems \cite{SC08,SCH04}. In spite of the widespread adoption of the rain event concept, a proper non arbitrary statistical 
definition of rain event is still an open issue. The word ``event'' has been used to indicate different things in different contexts, and a great variety of techniques 
have been used for its definition. The most common criterion to define a rain event is the Minimum Intra event 
Time (MIT): the MIT value separating two events ranges from 15 min to 24 hours \cite{DD08}. In many cases, the 
choice of a particular MIT is not related to observed dynamical properties of the rainfall phenomenon but to 
a particular time scale of the hydrological process considered: e.g. Lloyd et al. \cite{LCR90} use 3 h as MIT because it is 
the estimated time for the canopy to dry, Bracken et al. \cite{Betal08} use MIT=12 h so that the ground could dry 
between rain off events, while Aryal et al.  \cite{Ary07} use 8 h as MIT because it is the estimated time for 
highway surface depression to dry. 

Recently, an alternative definition of rain event \cite{PHC01,PC02,PC06}  
has gained some recognition in hydrological literature \cite{ARFS03,AEG03,DR04,TEL04,BPDL06,TEL07,MAR08}. According to this new proposal a 
rain event is identified by the occurrences of consecutive adjacent wet time intervals 
(the time intervals being equal in duration to the instrument time resolution). 
We will refer to this method as the Adjacent Wet Intervals (AWI) method. The adoption of this definition of 
rain event plays a central role in what is the main claim of Peters et al. \cite{PHC01,PC02,PC06}: a dynamical ``equivalence" between
the occurrence of rain events and that of avalanches in sand pile models. In this case rain would have the same dynamical properties 
of self-organizing systems \cite{BTW87}. These dynamical properties are commonly referred to, by the scientific community, 
with the term Self Organized Criticality (SOC). 

The MIT and AWI methods appear to be very different: the MIT method looks for the lack of rain for a suitable interval of time, 
while the AWI method looks for time intervals with a ``continuous'' presence of rain. We will compare both the MIT and AWI methods 
and provide theoretical arguments to show that the two methods are ``equivalent'' and therefore subject to the same arbitrariness: the 
choice of a minimum rainless interval in the MIT case, and a particular time resolution of the instrument in the AWI case.

As a way out from the issue of arbitrariness, we propose here to focus on the dynamical properties of the rainfall 
phenomenon and see if a particular time scale exists which could be adopted as a physically meaningful (and thus not arbitrary) 
rainless period (time resolution) in the MIT (AWI) case.
There are two kinds of precipitation regimes: \textit{stratiform} and \textit{convective} \cite{HOUBOOK,HOU97}. 
The words \textit{stratiform} and \textit{convective} do not refer to the process of cloud formation but to two distinct 
dynamical mechanisms of drop formation 1) \textit{stratiform}: drops are formed via condensation which is predominant in the 
case of small ($<$2m/s) updraft velocity, and 2) \textit{convective}: drops are formed via coalescence which is predominant in 
the case of large ($>$2m/s) updraft velocity. The occurrence of convective precipitation 
at mid latitudes is restricted to thunderstorm activity (isolated summer thunderstorm or Mesoscale Convective Systems (MCS)), while both 
types of precipitation occur between the tropics in spite of the fact that convection is the only mechanism of cloud formation, e.g.: \cite{HOU97}. 

Since physics laws are independent from the particular location on Earth, one expects that two stratiform (convective) events must have some 
common statistical properties since the rain drops were generated by the same physical mechanism. Note that the similarities we are referring to  
are not similarity in the duration or depth of showers, although they might exists, but similiraties in the statistical properties of inter drop 
time intervals and drop diameters. In other words, it is not ``how much'' or`` how long'' it rains which is similar 
(since orographic condition and other particular meteorological factors may be relevant) but ``how'' it rains. Newton's 
law $\vec{f}$$=$$m$$\vec{a}$ tells us that body's acceralation is always proportional to the applied force, but the trajectory of 
a body will depend on the the peculiar force applied. Eventual similar statistical properties of inter drop time intervals and drop diameters 
would be more easily detected by direct study rather than through the observation of rainfall rates at a given time 
resolution where things are complicated by the integration of the cube of the drop diameters over an arbitrary time interval (time resolution). 
Ignaccolo et al. \cite{IDMB09} have shown that invariant statistical properties can be obtained for the inter drop time intervals 
and drop diameters in the case of stratiform precipitation. We will use these properties to propose a non arbitrary definition 
of rain event for the case of stratiform rain, and to describe the internal variability of rain events.

In this manuscript, we use data from a Joss-Waldvogel (JW) disdrometer located at Chilbolton, UK for which the precipitation is almost 
exclusively of stratiform kind. Our data span a period 
of $\sim$2 years and are divided in 8 separate intervals of continuous observation (all data details are in the 
Appendix \ref{ourdata}). In Section \ref{droplike}, we review the properties of the sequences of inter drop time intervals 
and drop diameters which are relevant to our purposes. In Section~\ref{arbre}, we compare the MIT- and AWI-based definitions of event 
and show they are  ``equivalent''. A non arbitrary criterion to define rain events is introduced in Section \ref{ourdef}.  
Moreover, we describe the intra event dynamical variability using the concepts of active and quiescent phases.
Finally, we draw our conclusions in Section \ref{conclusion}.
 

\section{Drop-like nature of rain and the statistical properties of stratiform rain}\label{droplike}
Rainfall is discrete process being a collection of drops. Considering a small portion of 
Earth's surface, such as the collecting area of disdrometer or a rain gauge, rain can be described mathematically as a sequence of couples $(d_{j}$,$\tau_{j})$, 
$j$$=$1,2,3,$\dots$. The symbol $d_{j}$ indicates the diameter of the j-th drop, while $\tau_{j}$ is 
the time interval (inter drop time interval) between the j-th and $($j$+$1$)$-th drop. 
However, every instrument designed to detect rainfall has a time resolution \mbox{$\Delta$$>$$0$}: the 
instrument integration time. As a consequence, all inter drop time intervals of duration
\mbox{$\tau$$<$$\Delta$} are lost, and all inter drop time intervals of duration \mbox{$\tau$$>$$\Delta$} are detected  
as dry intervals, drought, of duration $\left[\tau / \Delta \right]$$\times$$\Delta$ or $($$\left[\tau / \Delta \right]$$-$$1$$)$$\times$$\Delta$
($\left[.\right]$ indicates the integer part) \cite{IDMB09}.

Let us consider as example $\Delta$$=$1s. All inter drop time intervals $\tau$$<$1s cannot be detected, and the subsequent drops will ``wet'' the time intervals of 
duration $\Delta$$=$1s containing them. An inter drop time interval $\tau$$=$1.3 s has 30\% probability of resulting in a single dry interval (drought 
of duration 1 s), and 70\% probability of not being detected with the subsequent drop wetting the time interval containing it. 
A value of $\tau$$=$3.6 s will generate 3 consecutive dry intervals (drought of 3 s) in 60\% of the cases, and 2 consecutive dry 
intervals (drought 2 s) in the remaining 40\% of the cases. The above limitations are common 
to the great majority of rainfall measuring instruments. The only exceptions are instruments with high temporal resolutions ($\sim$1 ms) 
such as the video disdrometer \cite{LG98,LG06,KRUG02}. Thus, given a rainfall time series one is able to calculate, in most 
cases, only the probability $P(l\Delta)$ of observing $l$ consecutive dry intervals at resolution 
$\Delta$. A better statistical accuracy is obtained using the ``survival'' probability  $\bar{P}(l\Delta)$: the probability 
of observing $l$ or more consecutive dry intervals at resolution $\Delta$.

Neverthenless, it is possible to investigate the statistical properties of the inter drop intervals also using data with 
a time resolution much larger than the one offered by video disdrometer or particle spectrometer ($\Delta$$\gg$1ms) such 
as the time resolution of our data $\Delta$$=$10 s. In fact, it is possible to derive a formula describing the 
dependance of the probablity $P(l\Delta)$  on the probability density function $\psi(\tau)$ of having 
an inter drop time interval $\tau$  \cite{IDMB09}. In the limit $\tau$$\gg$$\Delta$ ($l$$\gg$1), this dependence is rather simple:
\begin{equation}\label{nicesurv} 
P(l\Delta) \; \mathop{\propto}_{\tau\gg\Delta}\; \psi(l\Delta) \;\;\Leftrightarrow\;\; \bar{P}(l\Delta)\;\mathop{\propto}_{\tau\gg\Delta}\; \Psi(l\Delta),
\end{equation}
where the symbol $\Psi(\tau)$ indicates the probability of having an inter drop time interval $\geq$$\tau$ (survival probability).
\subsection{Statistical properties of stratifrom rain}\label{stratiproperties}
In the following, we briefly report some  statistical properties of the sequence of couples $(d_{j}$,$\tau_{j})$ in the case of stratiform rain 
derived in \cite{IDMB09} to which we refer the reader for a detailed discussion. The statistical properties relevant for 
the purposes of the present manuscript are:

1) Drop diameters and inter drop time intervals are not independent as ``large'' inter drop time intervals separate drops of 
``small'' diameter. This conclusion was derived from the observation of the conditional frequencies for a drop diameter given a specific value of 
the preceeding inter drop time intervals. In particular, inter drop time intervals $\gtrsim$10 s are followed in $>$95\% of the cases by a drop which 
diameter is $\leq$0.6 mm.

2) Large inter drop time intervals ($\tau$$>$10s) are preceded/followed either by another large inter drop time interval or by a sequence of few small ($\tau$$<$10s) 
inter drop time intervals. In this second case, the occurrence of 5 or less drops in the 10 seconds (a rate 0.5 drops per s) following a 
large inter drop time interval is higly probable ($\geq$95\%).

\begin{figure}[h]
\includegraphics[angle=0,height=180mm,width=1.0\linewidth]{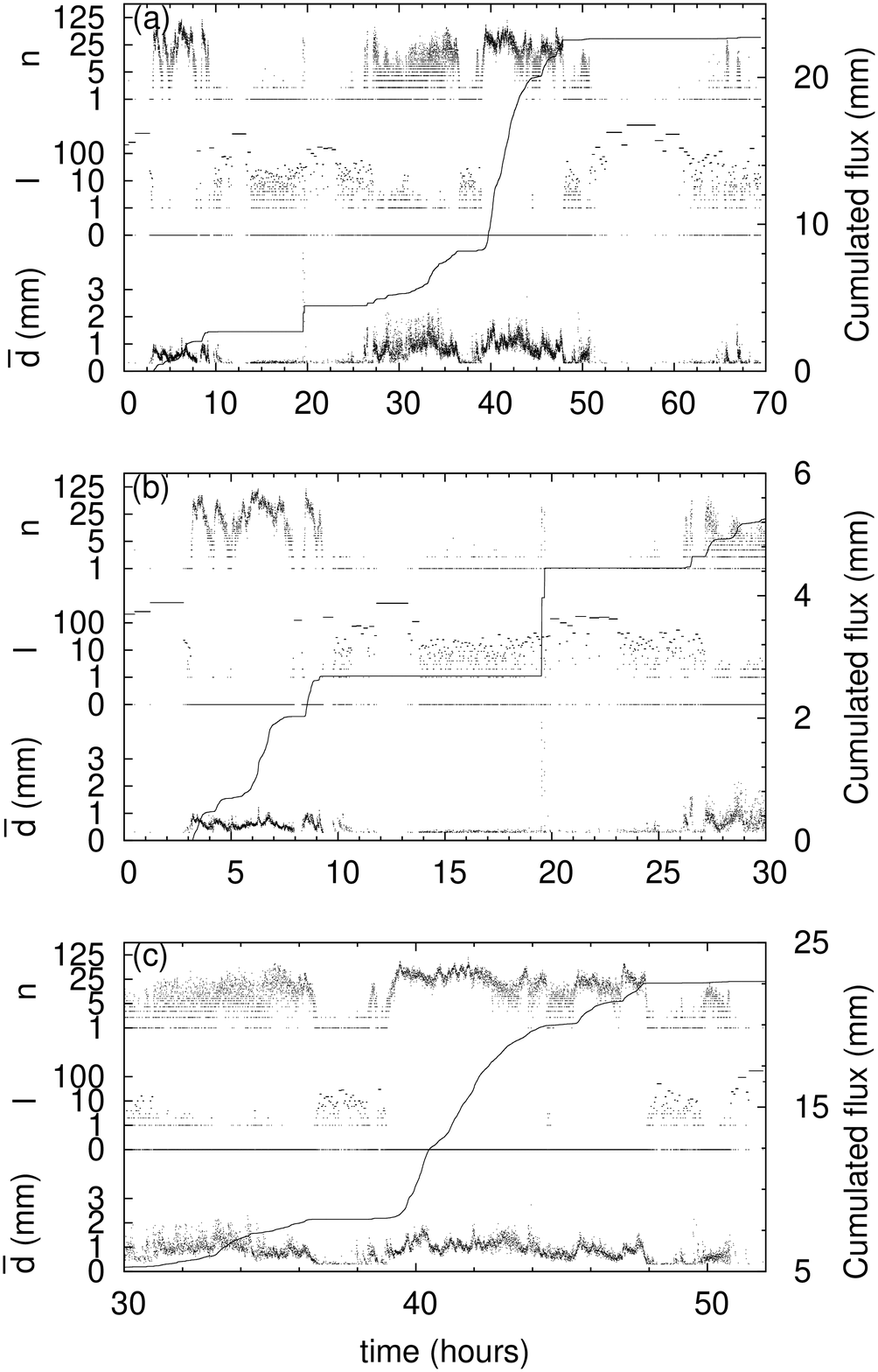}
\caption{A $\sim$3 days extract from our data set. For each time interval 
$\Delta$$=$10 s (instrument resolution), we plot the number $n$ of drops fallen, 
and the average drop diameter $\bar{d}$ in the interval. Moreover, we plot the 
number $l$ of consecutive dry intervals (droughts of duration $l\Delta$). A drought of duration 100$\Delta$ 
is represented by a horizontal bar of 100 data points at the ``100'' level. Each 
wet interval (interval occupied at least by one drop) is represented as a drought of 
null duration: ``0'' level. Finally the continuous line represents the cumulated 
flux. Panel (a) shows the above mentioned quantities for the entire duration of 
the extract (70 hours), panel (b) zooms on the first 30 hours, and panel (c) zooms on 
the 30 to 52 hours range. Note how the properties 1) and 2) described in 
Section~\ref{stratiproperties} create extended regions in which precipitation 
occurs without a detectable increase of the cumulated flux: ``quiescent'' regions. 
E.g. consider the time intervals 10-19.5 h and 20-26 h in panel (b), or the 
time intervals 26.5-28 h and 48-50 h in panel (c).}\label{figure1}
\end{figure}

\begin{figure}[h]
\includegraphics[angle=0,height=180mm,width=1.0\linewidth]{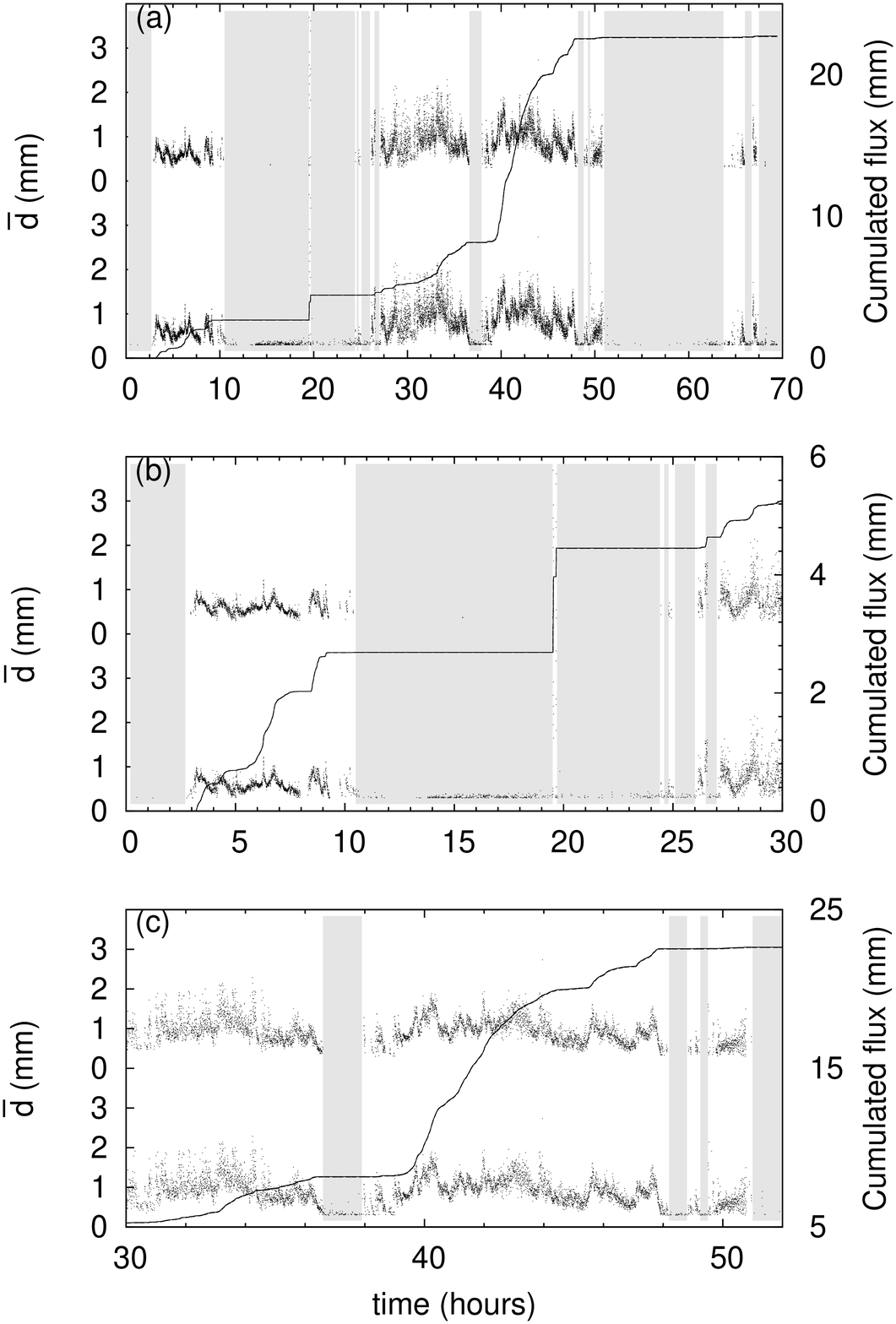}
\caption{The average diameter $\bar{d}$ in a time interval of
duration $\Delta$$=$10s and the cumulated flux before (bottom half
dots and solid line) and after (upper half dots and dashed line)
removing the quiescent phases as a function of time. The gray shaded areas provide 
visual aid highlighting quiescent regions. Panel (a), the entire
70 hours of observation, panel(b) zooms in to the first 30 hours, and panel (c) 
zooms on the 30 to 52 hours range.}
\label{figure2}
\end{figure}

3) Two different dynamical regimes of the rainfall phenomenon can be observed. These are the quiescent and active phases. 
The contribution to the total cumulated flux of quiescent phases is negligible (they are time intervals of sparse precipitation), while non quiescent regions carry 
the bulk of the precipitated volume, hence they are ``active''. Fig.~\ref{figure1} is a $\sim$3 days extract from our Chilbolton data showing the alternance of 
quiescent and active phases. Note that the quiescent phase is not just what is commonly referred as a dry spell or drought (e.g.: 5
days without precipitation are obviously a quiescent time interval) but it is a more dynamically reach concept. For example, consider the time interval 15-25 h as 
depicted in panel(b) of Fig.~\ref{figure1}. If one hour resolution is adopted, the time interval 15-25  h will result in a sequence of 
10 consecutive wet intervals which could be considered as an event. However, panel (b) of Fig.~\ref{figure1} shows that such a sequence of wet time intervals 
is constituted by two quiescent phases (almost null flux increase) and an active phase in between (19-20 h) in which a sharp increase of 
the cumulated flux occurs . The presence of quiescence and active phases is a direct consequence of properties 1) and 2). 
Property 2) indicates the occurence of time intervals of sparse precipitation since large inter drop time intervals are preceded and followed by the 
occurrence of small rainfall rates ( $\leq$0.5 per second). While property 1) indicates the arrival of drops of small diameter ($d$$\leq$0.6 mm) after large 
inter drop time intervals genarating a negligible cumulated flux. As a 
matter of fact, one can use property 1) and 2) to be label as quiescent all regions with a drop arrival rate $\leq$0.5 per second, and with an average drop diameter $\leq$0.6 
mm (see \cite{IDMB09} for the detail implementation of this procedure). 
The results of this filtering process is shown in Fig.~\ref{figure2}. We see how the described filtering procedure correctly indentifies the 
regions of null or almost null increase of the cumulated flux: quiescent phases. 

4) The probability density function of inter drop time intervals $\psi(\tau)$ has an inverse power law regime in a certain range. We will 
define the borders of this interval in the next Section. Note that using 1), 2) and 3), we can conclude that the inverse power law regime for the frequency 
of inter drop time intervals is a dynamical charateristc of the quiescent phase.


\section{Arbitrary definitions of rain event}\label{arbre}
Even if a physical identification of rain events based on meteorological considerations accounting
for atmospheric circulation and storm dynamics is preferable \cite{LCR90}, 
a statistical identification based on threshold criteria is widely used in hydrologic Literature for its simplicity in applications. 
\subsection{The Minimum Inter event Time (MIT)}\label{mitsum}
According to the MIT criterion, a rain event is the time interval in between the occurrence of a dry 
period equal or larger in duration to a specific threshold: the minimum intra event time or MIT. 
The range of MIT used in Literature is [15 min, 24 h]. The choice of a particular MIT value 
influences the number of rain events, the distributions of event duration, depth, and rain rate. 
Dunkerley \cite{DD08} calculates empirical formulas linking the number of rain events, the geometric mean event duration, the geometric mean event volume, 
the geometric mean event rain rate, and the geometric mean inter event time to the MIT values. 
Hereby, we plot in Figs.~\ref{figure3} and \ref{figure4} the survival probabilities $\bar{P}^{\textrm{\tiny{MIT}}}(\theta)$, the probability of observing 
a rain event of duration $\geq$$\theta$, and $\bar{P}^{\textrm{\tiny{MIT}}}(w)$, the probability of observing a rain event depth $\geq$$w$ for different 
MIT values $T$. The shapes of the probabilities $\bar{P}^{\textrm{\tiny{MIT}}}(\theta)$ and $\bar{P}^{\textrm{\tiny{MIT}}}(w)$ heavily depend on the 
particular value $T$.

\subsection{Adjacent Wet Intervals (AWI)}\label{awire}
According to the proposal of Peters et al. \cite{PHC01,PC02,PC06}, a rain event is a sequence of consecutive ``wet'' (precipitation occurred) time 
intervals of duration $\Delta$: the instrument resolution. Note however that given a rainfall time series with time resolution $\Delta$, one 
can build the corresponding time series for resolutions which are multiple of the time resolution (2$\Delta$, 3$\Delta$, $\dots$). 
As for the MIT case, we plot in Figs.~\ref{figure3} and \ref{figure4} the survival probabilities 
$\bar{P}^{\textrm{\tiny{AWI}}}(\theta)$ and $\bar{P}^{\textrm{\tiny{AWI}}}(w)$ relative to the rain event duration and 
rain event depth. Also in this case, we notice a marked dependence on the time resolution $\Delta$. 
The dependence on the time resolution $\Delta$ observed in Figs.~\ref{figure3} and \ref{figure4} for 
the survival probabilities $\bar{P}^{\textrm{\tiny{AWI}}}(\theta)$ and $\bar{P}^{\textrm{\tiny{AWI}}}(w)$ 
casts doubts on the claims that a proper theoretical framework for 
the rainfall phenomenon is the Self Organized Criticality \cite{PC06}. 
In fact the claim is based on the observed inverse power law for the survival 
probability  of observing a depth of duration $w$, in the case $\Delta$$=$1min. This behavior 
is typical of self organizing systems such as Earth's crust under stress and deformation 
which is why Peters et al. refer to the inverse power law behavior of rain event depths as rain's Gutenberg-Richter law \cite{GR54}. 
Fig.~\ref{figure4} shows that one could fit an inverse power law function to the survival probability 
$\bar{P}^{\textrm{\tiny{AWI}}}(w)$ relative to the resolutions $\Delta$$=$10 s and $\Delta$$=$1 min and obtain (in the range $[$0.001,1$]$ mm) two rather 
different exponents : $\mu$$=$0.48 and $\mu$$=$0.33 respectively. The exponents of the probability density function $\psi(\tau)$ are obtained adding 
1 to the exponent of the survival probability $\Psi(\tau)$. Moreover, for larger time resolutions the inverse 
power law character of the survival probability $\bar{P}^{\textrm{\tiny{AWI}}}(w)$ becomes less evident and 
the eventual exponent gets closer to zero: $\mu$$=$0.14, $\mu$$=$0.09, and  $\mu$$=$0.08 for $\Delta$$=$10 min, 1 h, and 3 h (the range $[$0.001,0.1$]$mm 
was used for these three last fits).
\subsection{Equivalence between the MIT- and AWI-defined event}\label{miteqawi}
\begin{figure}[h]
\includegraphics[angle=-90,width=1.0\linewidth]{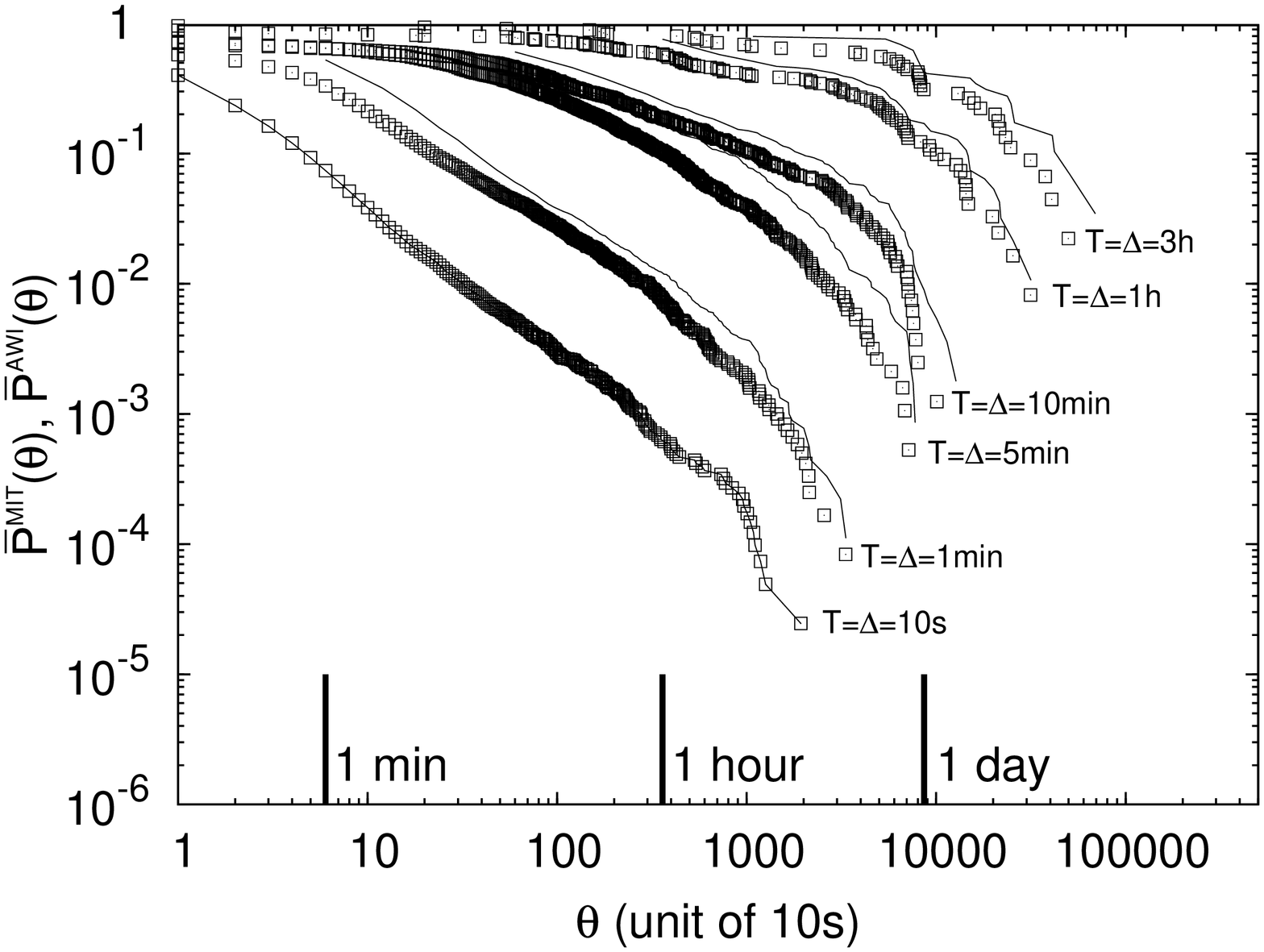}
\caption{The survival probabilities $\bar{P}^{\textrm{\tiny{MIT}}}(\theta)$  (squares) and $\bar{P}^{\textrm{\tiny{AWI}}}(\theta)$ (solid lines) for 
different values of $T$ (MIT method) and time interval $\Delta$ (AWI method) respectively. The time interval of continuous observation used for 
this figure is the one from 01/24/2004 to 05/11/2004.} \label{figure3}
\end{figure}

\begin{figure}[h]
\includegraphics[angle=-90,width=1.0\linewidth]{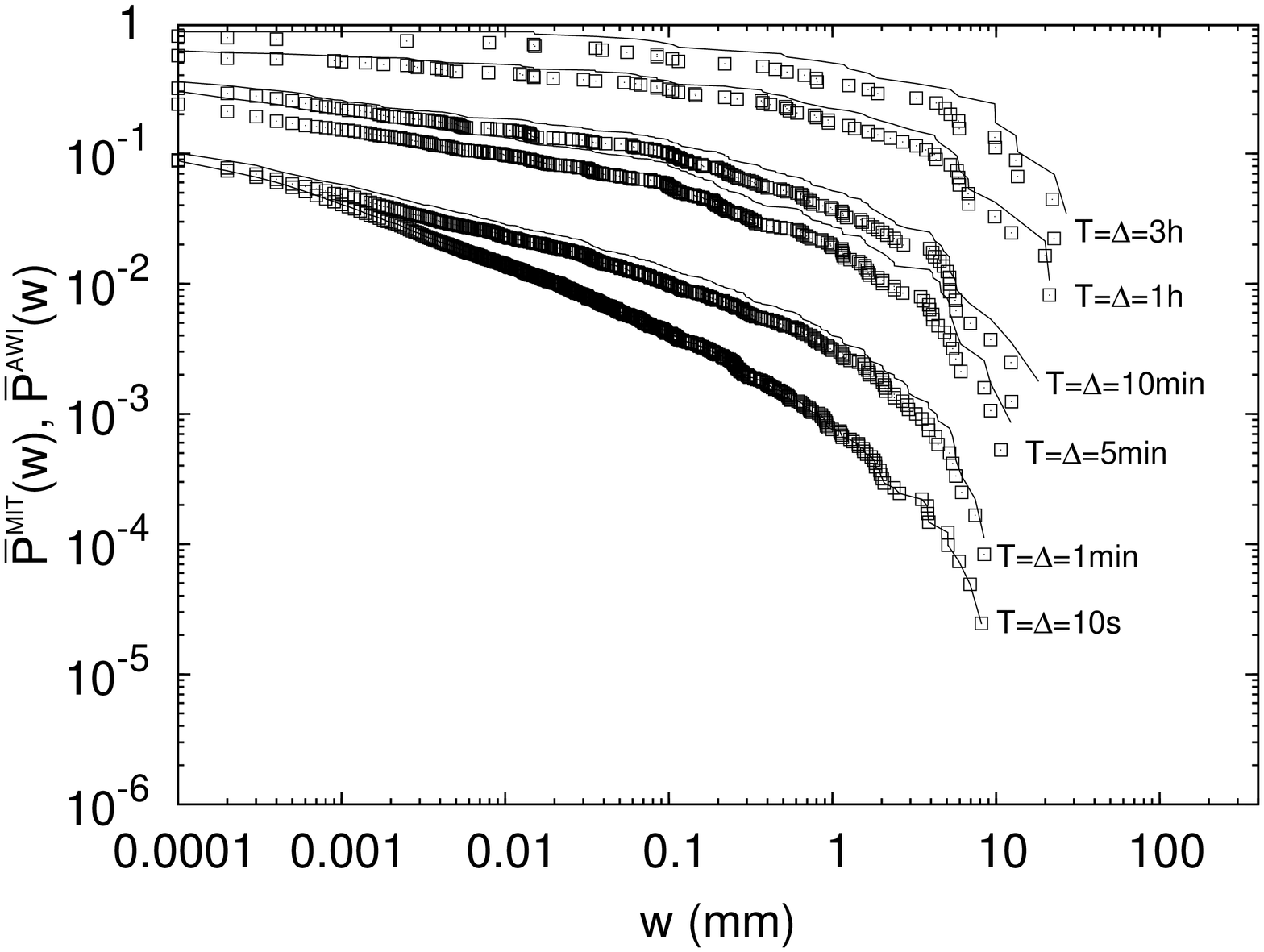}
\caption{The survival probabilities $\bar{P}^{\textrm{\tiny{MIT}}}(w)$  (squares) and $\bar{P}^{\textrm{\tiny{AWI}}}(w)$ (solid lines) for 
different values of $T$ (MIT method) and time interval $\Delta$ (AWI method) respectively. The time interval of continuous observation used for 
this figure is the one from 01/24/2004 to 05/11/2004.} \label{figure4}
\end{figure}

\begin{figure}[h]
\includegraphics[angle=-90,width=1.0\linewidth]{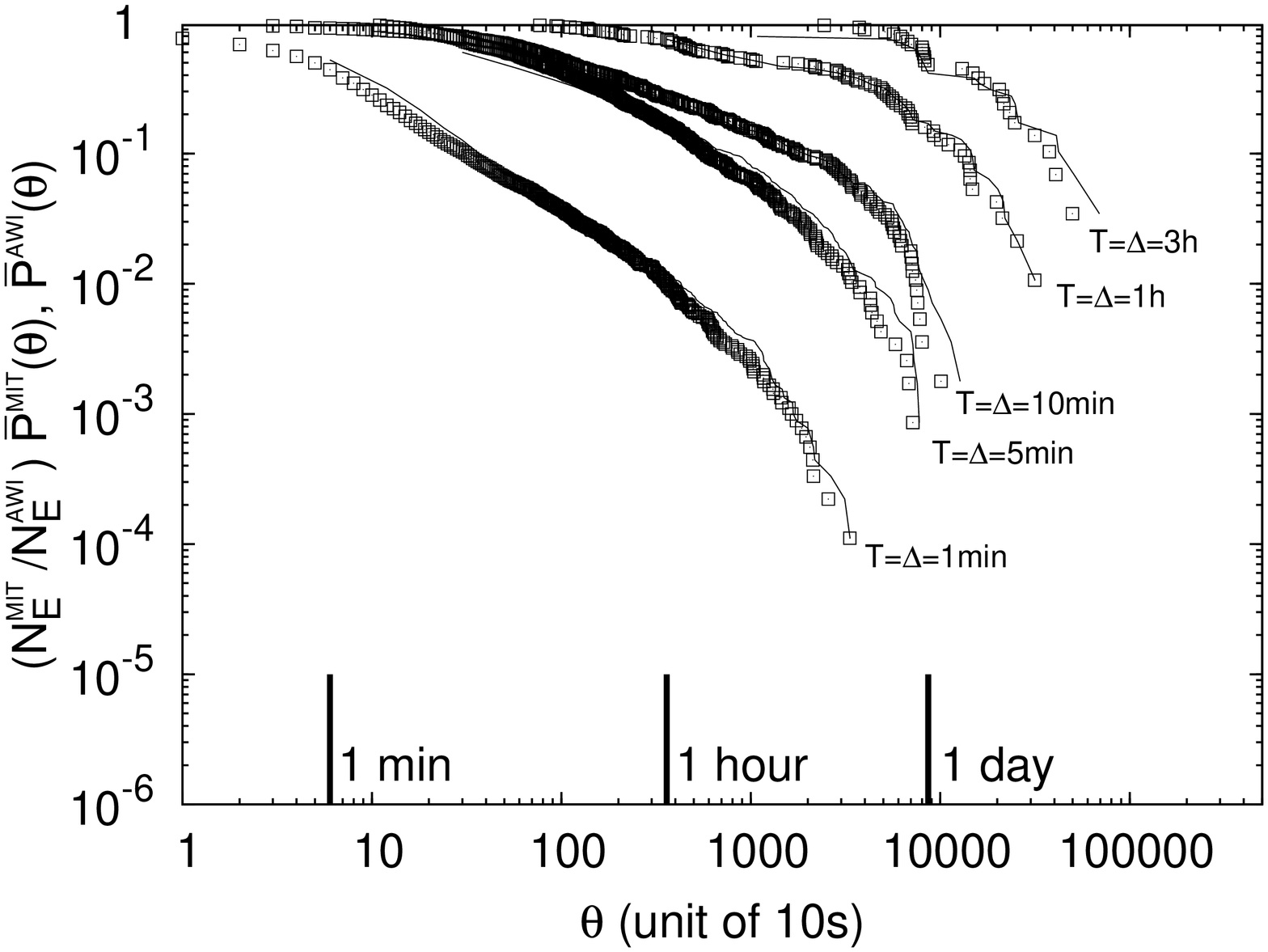}
\caption{The survival probability $\bar{P}^{\textrm{\tiny{AWI}}}(\theta)$  (squares) and the function $(N^{\textrm{\tiny{MIT}}}_{E}$/$N^{\textrm{\tiny{AWI}}}_{E})$
$\bar{P}^{\textrm{\tiny{MIT}}}(\theta)$ (solid lines) for different values of $T$ (MIT method) and time interval $\Delta$ (AWI method) respectively. The time 
interval of continuous observation used for this figure is the one from 01/24/2004 to 05/11/2004.} \label{figure5}
\end{figure}

\begin{figure}[h]
\includegraphics[angle=-90,width=1.0\linewidth]{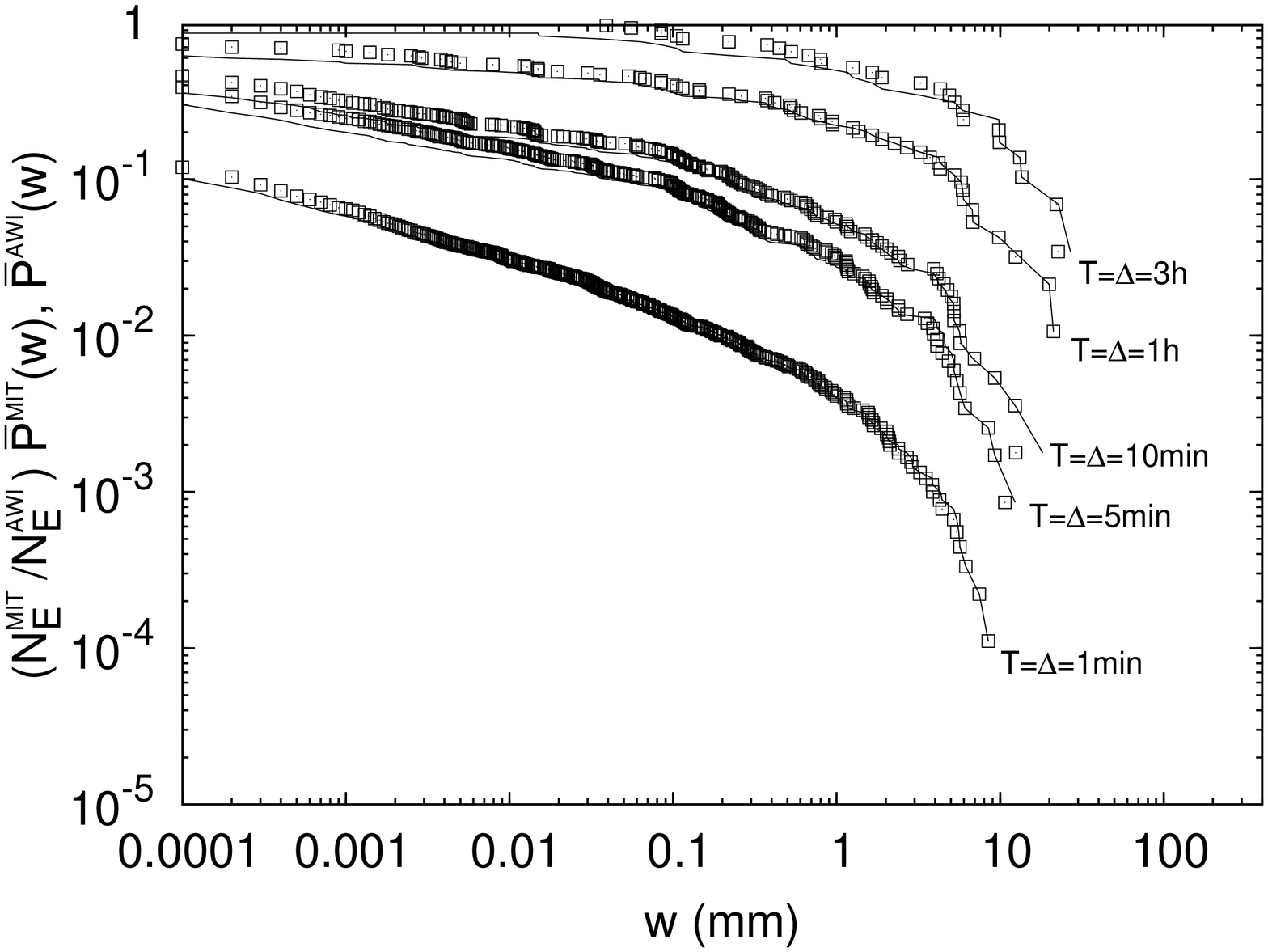}
\caption{The survival probability $\bar{P}^{\textrm{\tiny{AWI}}}(w)$  (squares) and the function $(N^{\textrm{\tiny{MIT}}}_{E}$/$N^{\textrm{\tiny{AWI}}}_{E})$
$\bar{P}^{\textrm{\tiny{MIT}}}(w)$ (solid lines) for different values of $T$ (MIT method) and time interval $\Delta$ (AWI method) respectively. The time 
interval of continuous observation used for this figure is the one from 01/24/2004 to 05/11/2004.} \label{figure6}
\end{figure}

Hereby, we show that the definitions of rain event based on the MIT and the AWI methods are ``equivalent''.
Let us consider a MIT of duration $T$, a rain event is a series of drops arrivals with inter drop time intervals 
which never exceeds  $T$. Such a sequence of drops would be perceived as part of a rain event also using the AWI 
method provided that the time resolution adopted in the rainfall record is $\Delta$$=$$T$. In this case all inter drop 
time intervals $\tau$$\leq$$T$ are lost (see arguments in Section~\ref{droplike}), and all the time intervals of duration $\Delta$ 
covering a MIT$=$$T$ defined event are wet. However, a AWI-defined rain event
via a time resolution $\Delta$$=$$T$  may contain inter drop time intervals exceeding $T$ up to a value 2$T$ (see arguments in Section~\ref{droplike}). 
As a consequence, a AWI event may be split in two or more parts if the MIT criterion 
is adopted. Therefore, AWI-defined rain events have a statistically longer duration and a larger depth than 
the rain events obtained with the corresponding MIT as confirmed by Figs.~\ref{figure3} and \ref{figure4}. 
The only exception to this rule is the case in 
which the values of both $T$ and $\Delta$ are equal to the instrument time resolution $\Delta_{res}$. In this case the MIT 
criterion can not split AWI-defined events based on the occurrence of inter drop time intervals in the range 
$[$$\Delta_{res}$,$2\Delta_{res}$$]$ because these intervals have not been detected by the instrument: Figs.~\ref{figure3} and \ref{figure4}, 
case $T$$=$$\Delta$$=$$\Delta_{res}$$=$10s. Finally, these figures also show that when $T$$=$$\Delta$ the survival probabilities for both the MIT and AWI cases have similar 
features so that the MIT and AWI criteria are somewhat ``equivalent''. To quantify this equivalence when $T$$=$$\Delta$, we consider the following variables: 
the number of MIT-defined (AWI-defined) events with a duration $\geq$$\theta$ $N^{\textrm{\tiny{MIT}}}_{>\theta}$ ($N^{\textrm{\tiny{AWI}}}_{>\theta}$), and 
the number of MIT-defined (AWI-defined) events with a depth $\geq$$w$ $N^{\textrm{\tiny{MIT}}}_{>w}$ ($N^{\textrm{\tiny{AWI}}}_{>w}$). 
The two couples of numbers are related as follows
\begin{equation}
\label{twonumbers1}
\left\{ \begin{array}{l}
N^{\textrm{\tiny{AWI}}}_{>\theta} = N^{\textrm{\tiny{MIT}}}_{>\theta}+C^{\textrm{\tiny{MIT}}}_{<\theta}-C^{\textrm{\tiny{MIT}}}_{>\theta}\\
N^{\textrm{\tiny{AWI}}}_{>w} = N^{\textrm{\tiny{MIT}}}_{>w}+C^{\textrm{\tiny{MIT}}}_{<w}-C^{\textrm{\tiny{MIT}}}_{>w}
\end{array} .\right.
\end{equation}
The symbol $C^{\textrm{\tiny{MIT}}}_{<\theta}$ ($C^{\textrm{\tiny{MIT}}}_{<w}$) in Eq.~(\ref{twonumbers1}) is the number of 
events gained via connection of two or more MIT events which singularly have all a duration (depth) $<$$\theta$ ($<$$w$). Similarly, 
$C^{\textrm{\tiny{MIT}}}_{>\theta}$ ($C^{\textrm{\tiny{MIT}}}_{>w}$) is the number of events lost via connection of two or more MIT events 
which singularly had already a duration (depth) $\geq$$\theta$ ($\geq$$w$). The two relations in  Eq.~(\ref{twonumbers1}) 
can be transformed in relations among the corresponding survival probabilities if the r.h.s. and l.h.s of both relations 
are multiplied by the constant factor 
$N_{E}^{\textrm{\tiny{AWI}}}$/$(N_{E}^{\textrm{\tiny{MIT}}}$$N_{E}^{\textrm{\tiny{AWI}}})$, 
where $N_{E}^{\textrm{\tiny{MIT}}}$ ($N_{E}^{\textrm{\tiny{AWI}}}$) is 
the total number of MIT (AWI) events. After some algebra, we obtain 
\begin{equation}
\label{twonumbers2}
\left\{ \begin{array}{l}
\bar{P}^{\textrm{\tiny{AWI}}}(\theta) = \frac{N^{\textrm{\tiny{MIT}}}_{E}}{N^{\textrm{\tiny{AWI}}}_{E}} \; \left ( \bar{P}^{\textrm{\tiny{MIT}}}(\theta)+c^{\textrm{\tiny{MIT}}}_{<\theta}-c^{\textrm{\tiny{MIT}}}_{>\theta} \right ) \\
\bar{P}^{\textrm{\tiny{AWI}}}(w) = \frac{N^{\textrm{\tiny{MIT}}}_{E}}{N^{\textrm{\tiny{AWI}}}_{E}} \; \left ( \bar{P}^{\textrm{\tiny{MIT}}}(w)+c^{\textrm{\tiny{MIT}}}_{<w}-c^{\textrm{\tiny{MIT}}}_{>w} \right ) \\
\end{array} .\right.
\end{equation}
The symbol $c^{\textrm{\tiny{MIT}}}_{<\theta}$ ($c^{\textrm{\tiny{MIT}}}_{>\theta}$,$c^{\textrm{\tiny{MIT}}}_{>w}$, 
$c^{\textrm{\tiny{MIT}}}_{<w}$) is the number $C^{\textrm{\tiny{MIT}}}_{<\theta}$ ($C^{\textrm{\tiny{MIT}}}_{>\theta}$,
$C^{\textrm{\tiny{MIT}}}_{<w}$, $C^{\textrm{\tiny{MIT}}}_{>w}$) divided by the total number of 
MIT events $N^{\textrm{\tiny{MIT}}}_{E}$. Eq.~(\ref{twonumbers2}) shows that if we disregard the correction terms 
$c^{\textrm{\tiny{MIT}}}_{<\theta}$, $c^{\textrm{\tiny{MIT}}}_{>\theta}$, $c^{\textrm{\tiny{MIT}}}_{>w}$, and $c^{\textrm{\tiny{MIT}}}_{<w}$ 
the AWI survival probabilities $\bar{P}^{\textrm{\tiny{AWI}}}(\theta)$ and $\bar{P}^{\textrm{\tiny{AWI}}}(w)$ are proportional to the MIT 
survival probabilities $\bar{P}^{\textrm{\tiny{MIT}}}(\theta)$ and $\bar{P}^{\textrm{\tiny{MIT}}}(w)$. 
Thus in a log-log plot the two kind of distributions will differ by a constant factor which explains the ``similitude'' 
between MIT and AWI statistics observed in Figs.~\ref{figure3} and \ref{figure4}. To further test this conjecture we 
plot in Figs.~\ref{figure5} and \ref{figure6} the survival probability 
$\bar{P}^{\textrm{\tiny{AWI}}}(\theta)$ ($\bar{P}^{\textrm{\tiny{AWI}}}(w)$) with the rescaled survival 
probability $(N^{\textrm{\tiny{MIT}}}_{E}$/$N^{\textrm{\tiny{AWI}}}_{E})$$\bar{P}^{\textrm{\tiny{MIT}}}(\theta)$ 
($(N^{\textrm{\tiny{MIT}}}_{E}$/$N^{\textrm{\tiny{AWI}}}_{E})$$\bar{P}^{\textrm{\tiny{MIT}}}(w)$). Figs.~\ref{figure5} and \ref{figure6} 
confirm that the statistical properties of MIT- and AWI-defined events are 
``equivalent'': the correction terms in Eq.~(\ref{twonumbers2}) do no alter dramatically the shape of the survival probability functions.

\section{A non arbitrary definition of rain event}\label{ourdef}
In his review work \cite{DD08}, Dunkerley shows the great variability in the MIT value used to define an event, and also 
how in many cases events are defined but the criteria are not reported. Another issue related to the particular value of MIT adopted 
is the intra event ``variability'': namely the fact that an event is not characterized by the 
continuous presence of rain, but it is mostly rainless. About 75\% of the event duration is found to be rainless by \cite{PKAR07}. 
Hereby, we show how to address these important questions it is necessary to focus on the dynamical properties of the rainfall 
phenomenon.

In Section \ref{droplike} and Section \ref{stratiproperties}, we have discussed the properties of the sequence 
of the couples ($d_{j}$,$\tau_{j}$) and shown that two phases of rainfall phenomenon can be identified: 
a quiescent one and an active one. These findings and Eq.~(\ref{nicesurv}) allow us a dynamically interpretation 
of the survival probability  $\bar{P}(l\Delta)$

Let us consider the extreme values of the inter drop time intervals  ($\tau$$\rightarrow$$\infty$). 
The shape of the survival probability $\bar{P}(l\Delta)$ for extreme values ($l$$\gg$1) is a 
signature of the particular meteorological condition of the location in which data are gathered (e.g.: arid/non arid region).
Thus we refer to this region as the meteorological region. The problem 
of a ``proper'' choice of MIT (or time interval $\Delta$ in the AWI case) is related to where to locate the left border 
of the meteorological region. As we move from $+$$\infty$ to smaller values of inter drop time interval, we cross  
the border between the inter-storm dynamics (meteorological region) and the intra-storm dynamics (different showers belonging 
to the same synoptic event). Where to locate this border? We think that the emergence of a power law regime for the inter drop 
time intervals suggests a clear (non-arbitrary) ``dynamical'' choice. 

\begin{figure}[h]
\includegraphics[angle=-90,width=1.0\linewidth]{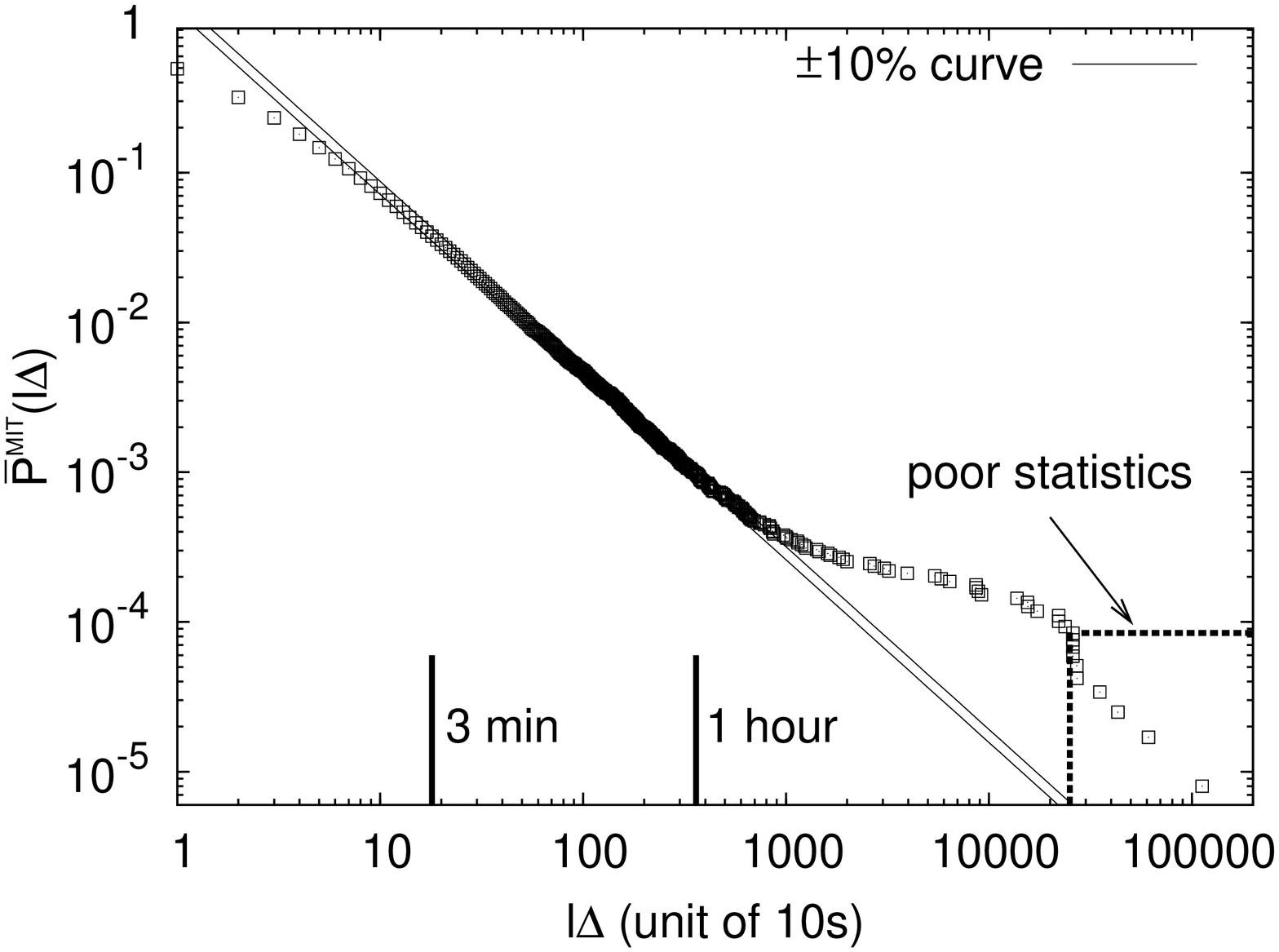}
\caption{Log-log plot of the survival probability
$\bar{P}(l\Delta)$  for the time interval of continuous
observation that goes from 04/01/2003 to 11/03/2003 (squares). The solid lines are indicate an error barof $\pm$10\% 
with respect to the inverse power law fitting curve in the range 3 min - 1 h. Finally the box in the bottom right corner 
(dashed lines) indicate he region of poor statistics, see Eq.~(\ref{cheby}).}\label{figure7}
\end{figure}

We adopt the following  operational method for locating this border. 
We fit with an inverse power law function the survival probability  $\bar{P}(l\Delta)$ in the range $[$$>$3 min, 1 h$]$, corresponding 
a range $[$$>$18, 360$]$ for $l$, given the resolution $\Delta$$=$$\Delta_{res}$$=$10 s of our data. The notation $>$3 min ($>$18) 
indicates that the left border is taken to be 3 min or more: we consider 3 min as the minimum value for which 
Eq.~(\ref{nicesurv}) is satisfied. However, there are cases for which the survival probability $\bar{P}(l\Delta)$ cannot be yet 
consider to have an inverse power law behavior (no approximate straight line in a log-log plot) for $l$$=$18 and larger 
values of $l$ ($\sim$30-40) are chosen since fit accuracy is dependent on a proper choice for the minimum value for which the 
inverse power law holds \cite{CLAU09}. Moreover, we choose $l$$=$360 (1 hour) for the right 
border of the fitting region, a value that is smaller of what is the expected (by visual inspection) limit of validity of the power law behavior. Once the fit is performed, 
we consider a $\pm$10\% difference from the value of the fitting curve as an indication of departure from the inverse power law regime.
Thus, we look at values of $l$$>$360 and seek the occurrence of at least 5 consecutive data points outside the 10\% error bar to declare as ``over'' 
the inverse power law regime. With this method the power law regime for the 
survival probability  $\bar{P}(l\Delta)$ relative to the time interval of continuous observation from 04/01/2003
to 11/03/2003 occurs in the $[$3 min, 2 h$]$ range as shown in Fig.~\ref{figure7}. Finally, we check if the calculation of the inverse 
power law range may be affected by lack of statistics. If $N$ droughts are present in a raifall record, which is the maximum value $l_{max}$ 
for the survival probability $\bar{P}(l\Delta)$ to be statistically reasonable? With $N$ samples the smallest ``probability'' 
detectable is 1/$N$. However one cannot define $l_{max}$ as the values for which the following equality $\bar{P}(l_{max}\Delta)$$=$1$/$$N$ holds: 
this is equivalent to say that all values of the survival probability are statistically sound. A more realistic and reasonable assumption can be done 
using the law of large numbers and in particular the Chebyshev's inequality 
\cite{GRIM01}. Let us consider an integer $K$$<$$N$ and the corresponding value $l_{K}$ for which the theoretical expected probability of having 
droughts equal or larger in value than $l_{k}$$\Delta$ is $K$$/$$N$.
According to the Chebyshev's inequality,
\begin{equation}\label{cheby}
\textrm{Pr}\left\{\bar{P}(l_{K}\Delta)>=2\frac{K}{N}\right\}\leq \frac{1}{K} \left(1-\frac{K}{N}\right). 
\end{equation}
The value  1/$K$(1-$K$/$N$)$\simeq$1/$K$ if $K$$\ll$$N$ is the maximum possible value for the probability (Pr) of observing, due to the finiteness 
of the sample ($N$$<$$\infty$), a value of the survival probability $\bar{P}(l_{K}\Delta)$ which is at least the double (2$K$/$N$) of the expected 
one. If we consider $K$$=$10$\ll$$N$ and the corresponding $l_{10}$, Eq.~(\ref{cheby}) says that the survival probability $\bar{P}(l_{10}\Delta)$ can be 
considered reasonably close to the expected theoretical value. In fact, there is probability $=$10\% that the observed value of $\bar{P}(l_{10}\Delta)$ 
is wrong by a factor 2 or more. Thus we set $l_{max}$$=$$l_{10}$: the dashed box in Fig.~\ref{figure7} indicates the region of ``poor'' statistics in the 
case of the time interval from 04/01/2003 to 11/03/2003. Table~1 reports the values of $l_{max}\Delta$ for all 8 time intervals 
of continuous observation available to us together with the number $N$ of droughts in the interval and the estimated range of validity of the inverse power law.
\begin{table}\label{table1}
\caption{The range of validity of the inverse power law regime, the number $N$ of droughts, and the values $l_{max}\Delta$ denoting 
the limit up to which the survival probability $\bar{P}(l\Delta)$ can be considered statistically sound  for all the 8 intervals of continuous observation 
in our data.}
\begin{tabular}{|c|c|c|c|c|}
\hline
Time interval of observation & Range & $N$ & $l_{max}$$\times$$\Delta$\\
\hline
04/01/03 -- 11/03/03 & 3 min - 2 h & 118417  & 69.4 h\\
\hline
11/05/03 -- 01/05/04 & 6 min - 1.5 h & 17879 & 22.2 h\\
\hline
01/08/04 -- 01/20/04 & 6 min - 32 min & 10875 & 32 min\\
\hline
01/24/04 -- 05/11/04 & 6 min - 4 h & 40781 & 24.2 h\\
\hline
05/14/04 -- 07/17/04 & 3 min - 2,77 h& 15519 & 2.77 h\\
\hline
07/19/04 -- 08/02/04 & 3 min - 1.61 h & 2857 & 1.61 h \\
\hline
08/04/04 -- 08/19/04 & 3 min - 53 min & 4612 & 53 min \\
\hline
12/10/04 -- 02/28/05 & 6 min - 1 h & 4401 & 2.55 h \\
\hline
\end{tabular}
\end{table}
Of the 8 time intervals of continuous observation only 3 (04/01/03 -- 11/03/03, 11/05/03 -- 01/05/04/, and 01/24/04 -- 05/11/04) have an 
accuracy (value of $l_{max}\Delta$) for the survival probability which is well above 20 h. The remaining intervals have an accuracy which 
does not permit to accurately locate the end of the power law regime. Considering the only 3 intervals with reasonable accuracy, we have 
for the limit of validity of the inverse power law regime the following values: 1.5, 2, and 4 h with an average value of 2.5 h.

\begin{figure}[h]
\includegraphics[angle=-90,width=1.0\linewidth]{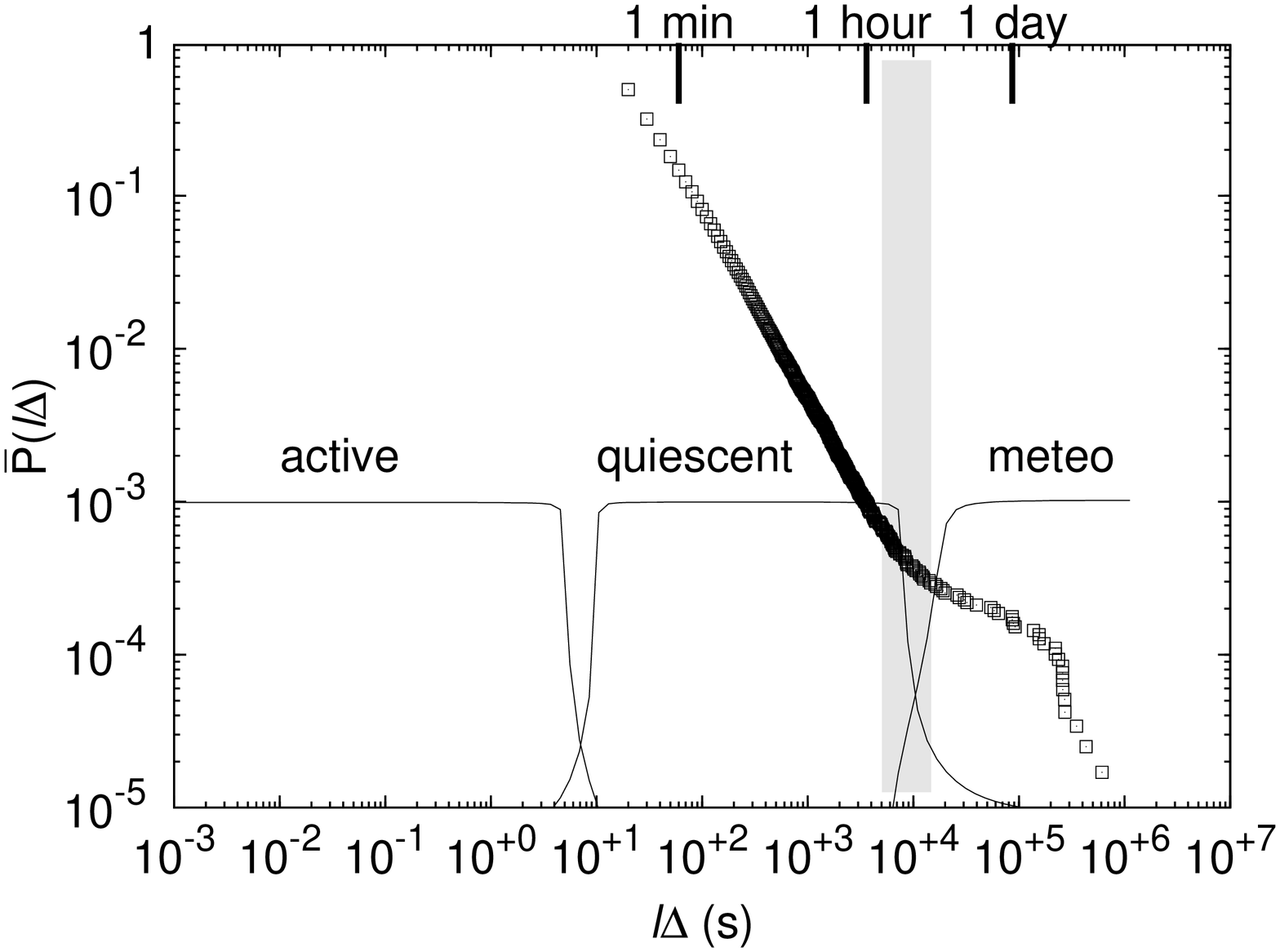}
\caption{Log-log plot of the survival probability
$\bar{P}(l\Delta)$  for the time interval of continuous
observation that goes from 04/01/2003 to 11/03/2003 (squares). The vertical lines indicate the proposed range for the ``proper'' choice of MIT. The other solid lines 
indicate qualitatively the probability (not normalized for better clarity) for 
an inter drop time interval $\tau$ to belong to each of the three dynamical phases of the rainfall phenomenon.} \label{figure8}
\end{figure}

Crossing, from right to left, the mark indicating the end of the inverse power law regime, we move from the meteorological region into the intra-storm quiescent region 
which from now on we will refer simply as the quiescent region. Moving to even smaller values of inter drop time intervals, 
we reach the intra-storm active region, from now on simply the active region. This transition cannot be observed with our data since the instrument 
resolution is $\Delta_{res}$$=$10 s. However, we know that active regions are characterized by a drop arrival rate $\gg$0.5 per second (Section 2.1). 
A drop arrival rate of 5 per second translates into an average inter drop time interval of 0.2 s. Moreover, we know (Section 2.1) that inter drop time intervals 
$\sim$10s belong to quiescent phases. Therefore, an educated guess for the time interval where to locate the transition from the quiescent phase 
into the active phase is $[$$\sim$1, $\sim$10$]$ s. This guess is confirmed by the plot of the probability density function $\psi(\tau)$ of inter drop time interval 
in Lavergnat et al. \cite{LG98,LG06} (obtained with an optic disdrometer which time resolution is $\Delta_{res}$$=$1 ms). The plot shows an inverse power law 
regime to emerge for inter drop time intervals $\tau$$\gtrsim$5 s. Fig.~\ref{figure8} shows the three dynamical regions 
together with the probability  $\bar{P}_{\Delta}(l\Delta)$ which is 
proportional, Eq.~(\ref{nicesurv}), to the survival probability of inter drop time intervals $\Psi(\tau)$. The vertical lines indicate the range $[$1.5, 4$]$ h. 
The other solid lines indicate ``qualitatively'' the probability for 
a given inter drop time interval to belong to a given dynamical region. A more ``quantitative'' description of these 
curves cannot be achieved with the data at our disposal: 1) for the transition between quiescent and meteorological region we need 
the synoptic information for the period of observation, 2) for the transition between the active and quiescent region 
we need a better instrument's time resolution.
\subsection{Intra event variability: active and quiescent phases}\label{intraedyn}
\begin{figure}[h]
\includegraphics[angle=-90,width=1.0\linewidth]{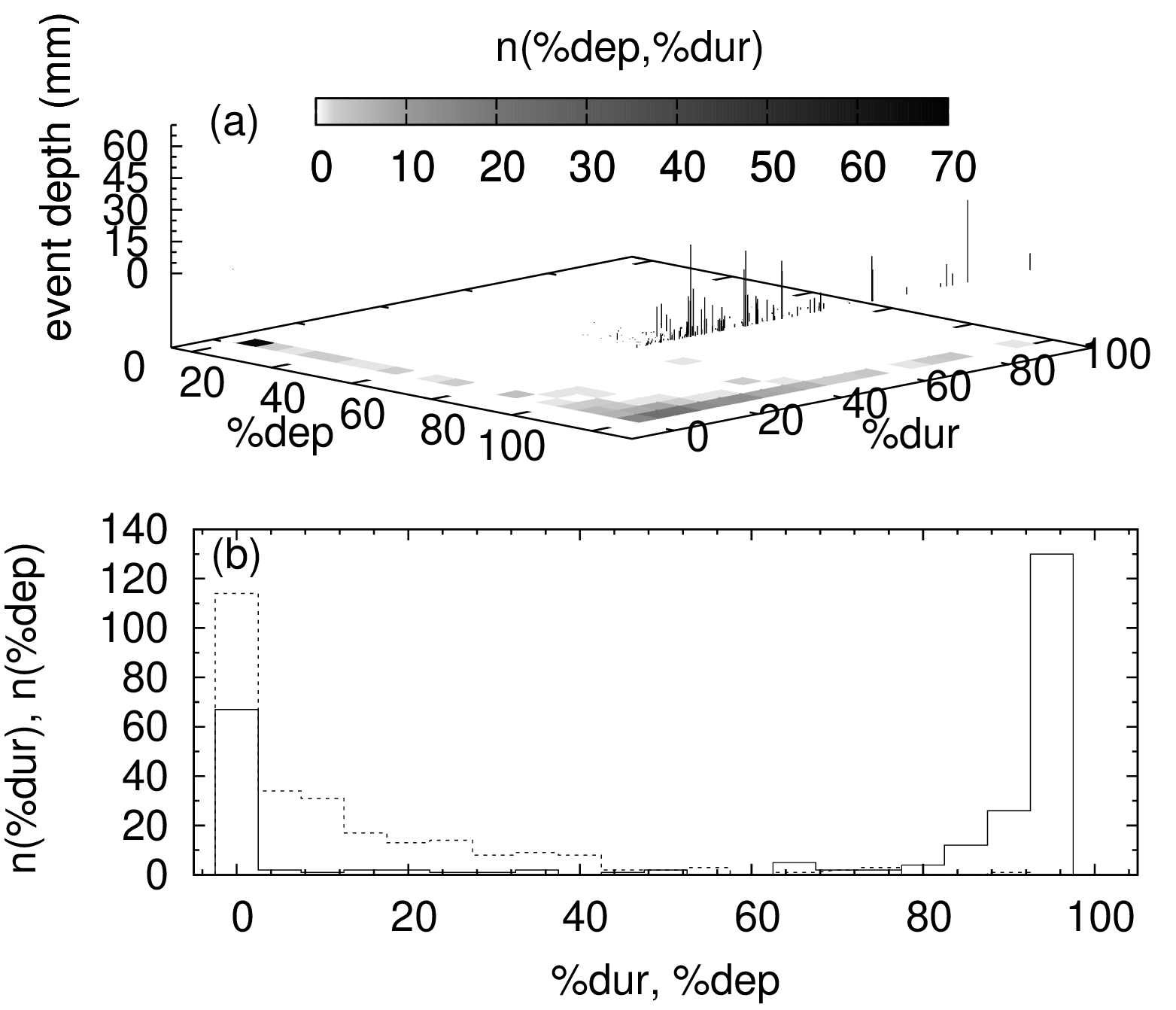}
\caption{panel (a). The number $n($\%dep,\%dur$)$ of events for which the active phases cover \%dur of the total duration and \%dep 
of the total depth (gray scale squares) calculated using a 5\% bin for both \%dep and \%dur. The depth of an event versus the percentage of 
duration \%dur and depth \%dep due to active phases (vertical lines). Panel(b) The plot of the number of events $n($\%dur$)$ with a given 
\%dur value (dashed line), and the number $n($\%dep$)$ of events with a given \%dep value (solid lines).}\label{figure9}
\end{figure}

Each event is composed by a non overlapping sequence of quiescent and active 
phases (Fig.~\ref{figure2}). To describe the intra event dynamical variability, we decide to calculate for each event which percentage of its duration and depth is due to 
active phases. We consider MIT$=$3h and all the 8 time intervals of continuous observation at our disposal (Appendix \ref{ourdata}) for 
a total of 262 events. The identification of quiescent and active phases inside each event is done using the same quiescent filter procedure 
used in \cite{IDMB09}. For each event, we calculate the percentage \%dur of the event duration which 
is covered by active phases, and the percentage \%dep of the event depth which is due to active phases. Then we calculate the number $n($\%dep,\%dur$)$ of events 
with a given \%dur and \%dep, and the number $n($\%dur$)$ ($n($\%dep$)$) of events with a given \%dur (\%dep). The results are shown in panels (a) and (b) of Fig.~\ref{figure9} 
respectively. The distribution of the number $n($\%dep,\%dur$)$ of events with a given \%dur and \%dep is roughly concentrated, panel (a), on a L-shaped region 
(the region of small values of \%dur and the region of large values of \%dep). This is confirmed by the plot of the numbers $n($\%dur$)$ and $n($\%dep$)$ 
of events with a given \%dur and \%dep, panel (b). Of the 262 events available, 62 have 0 active \%dep and \%dur: black square in (0,0) in panel (a), and left 
peak of the number $n($\%dep$)$ in panel (b). These events are produced by two different kinds of sequential ordering of inter drop time intervals $\tau$: 1) a 
$\tau$$>$MIT followed by a $\tau$$<$MIT followed by a $\tau$$>$MIT or 2) a $\tau$$>$MIT followed by a quiescent phase followed by a $\tau$$>$MIT. The first kind of 
sequence generates events constituted by a single drop, the second kind generates events composed by a single quiescent phase with negligible depth. 
The great majority of the remaining events (172 of 200) have an active \%dep$>$80\% and a \%dur which is mostly 
(151 of 172) less than 40\%, as indicated by the plot of $n($\%dur$)$ and $n($\%dep$)$ in panel (b). The left peak of $n($\%dur$)$, panel (b), is 
given by the 62 events with 0 active \%dep and \%dur, and by the events with \%dur$\leq$5\% and variable \%dep (one of the arms of the L-shaped region formed 
by the gray scale squares in panel (a)). Finally, the region of the (\%dur,\%dep) plane defined by the condition \%dep$\geq$80\% (right peak of $n($\%dep$)$ in panel (b)) 
is the region where all the ``relevant'' events (event depth $\gtrsim$1mm) are located as revealed by the plot of the event depth versus the \%dur and \%dep 
variables (vertical lines on panel (a)). Note how the depth of the event with \%dur$\leq$5\% and \%dep$\leq$80\% is so negligible that they are not resolved 
by the adopted scale for the z-axis in panel (a).
\section{Conclusions}\label{conclusion}
The question of ``what a rain event is'' has not been a simple one to answer in a statistical way. A plethora of methods have been proposed all of 
which affected by a degree of arbitrariness \cite{DD08}. We compare two methods for defining rain events: 
1) the MIT method, the most common \cite{DD08}, is based on defining a minimum rainless time interval $T$ separating two events. 
2) The AWI method which is based on the occurrence of consecutive wet intervals of equal duration $\Delta$. The statistical properties 
of the rain event duration and rain event depth are highly dependent from the particular value of $T$ (MIT-events) and $\Delta$ (AWI-events) chosen 
as shown by the plots of the survival probabilities (Figs.~\ref{figure3} and \ref{figure4} in Section 3.1 and 3.2). If for a given value of the parameter $T$ ($\Delta$) one could say 
that an inverse power law beahvior with scaling parameter $\mu$ exists for the survival probability of rain event duration and/or rain event depth, 
the same is not true for a different value of the parameter $T$ ($\Delta$): either the scaling parameter $\mu$ is sensibly different or there 
is no inverse power law regime. Consequently, claims made using survival probabilities (and/or probability density functions) for a particular value of $T$ ($\Delta$) should be 
considered void or doubtful at best. E.g., we doubt the conclusion reached by Peters et al. \cite{PC06}: inverse power law behavior for 
the probability of observing a rain event with depth $w$, in the case $\Delta$$=$1min, 
$\Rightarrow$ Gutenberg-Richter law for rainfall $\Rightarrow$ rainfall is an example of SOC. 

Moreover, we show that the MIT and the AWI methods are 
equivalent (if $T$$=$$\Delta$) since the corresponding survival probabilities for 
the rain event duration and depth have similar behavior (Eq.~(\ref{twonumbers1}), Eq.~(\ref{twonumbers2}), and Figs.~\ref{figure5} and \ref{figure6} 
of Section 3.3). This is so because the 
occurrence of adjacent wet intervals at time resolution $\Delta$ is related to the NOT occurrence of inter drop time intervals $>$$\Delta$ (Section~\ref{droplike}). 
It is worth noting that the equivalence between the MIT and AWI methods is not perceived as such in literature: the AWI method is presented as alternative 
method reflecting the ``burst''-like and scaling properties of the rainfall phenomenon, e.g. \cite{TEL04,TEL07}. This misconception is 
paired, in our opinion, with a lack of recognition of the drop-like nature of the rainfall phenomenon which result in the absence of detailed 
studies about the properties of the sequence of couples $(d_{j}$,$\tau_{j})$, drop diameter and inter drop time interval: \cite{IDMB09,LG98,LG06} being, 
to the best of our knowledge, the only exceptions.

Yet, the statistical properties of the sequence of inter drop time interval and drop diameter (properties 1), 2), 3), and 4) of Section~\ref{stratiproperties}) 
play a fundamental role also in answering the important question of what constitutes a rain event. We propose to use the ending time of the inverse power 
law regime present in the probability density function of inter drop time intervals $\psi(\tau)$ as the time $T$ (resolution $\Delta$) 
to adopt in the MIT (AWI) method. With the data at our disposal, 
we estimate (table 1) the ending time of the inverse power law regime to be in the range $[$1.5,4$]$h. Some considerations about this proposal are in order.
1) The proposal takes in account only the dynamics of the rainfall phenomenon and completely ignores, as it should be, any particular time scale 
relative to any other hydrological processes, e.g. catchment/runoff or canopy losses, which do not have any relevance for the occurrence 
of rain. This is not a limitation for the possible application of the proposed method. On the contrary, it is a strength: considering only the time 
scale of the particular hydrological application, while ignoring those relative to the rain and the possible complicated interplay of both, is not a recipe for 
success. 2) We consider this proposal valid only for \textit{stratiform} rain. The term \textit{stratiform} has to be intended as in \cite{HOU97}: 
updraft velocity in the clouds are typically less than 2 m/s 
leading to drop formation via condensation. This is the typical rain regime for Chilbolton, UK, and in general the predominant one at mid-latitudes. 
3) We deem this proposal to be non arbitrary because the inverse power law regime for the probability density function of inter drop time interval is a 
genuine dynamical properties of \textit{stratiform} rain. In fact the inverse power law regime is observed in all 8 time intervals 
of continuous observation, spanning almost 2 years, at 
Chilbolton (see \cite{IDMB09}) but also in other mid-latitude locations: Germany, France and Italy (\cite{IDMB09} and references therein). 
4) The criterion hereby discussed for a non arbitrary definition of rain event in  the case of \textit{stratiform} rain may have also some 
applicability in the case of  \textit{convective} rain, since \textit{stratiform} precipitation occurrs inside ``dying'' cell of convective storms \cite{HOU97}.

Finally in Section 4.1, we use the concepts of quiescent and active phases to describe 
the internal dynamical variability of a rain event. Rain events are mostly occupied by quiescent 
phases which precedes and/or follows ``showers'' responsible for the bulk of the event 
depth (active phases). There are many possible of ways of describing the rain events propeties and varability, e.g. 
quaterly volumes. A detailed comparison between our description of rain events in terms of quiescent and active phases 
and other metodology is beyond the scope of the present manuscript.

\appendix
\section{Chilbolton data}\label{ourdata}
The data used in this manuscript are collected at Chilbolton (UK) using a
Joss-Waldvogel impact disdrometer RD-69 \cite{JW67}, and provided by the British Atmospheric Data Centre. 
Precipitation at this location was monitored for a time interval of $\sim$2 years. However, some values are 
indicated as missing, and the data were carefully inspected to remove chunks were no value was reported. After 
this inspection, 8 different time intervals of continuous observation were identified: from 04/01/2003
to 11/03/2003, from 11/05/2003 to 01/05/2004, from 01/08/2004 to
01/20/2004, from 01/24/2004 to 05/11/2004, from 05/14/2004 to
07/17/2004, from 07/19/2004 to 08/02/2004, from 08/04/2004 to
08/19/2004, and from 12/10/2004 to 02/28/2005. The instrument has a collecting area $A$$=$50cm$^{2}$ and
provides every $\Delta$$=$10s the drop diameter count for 127 different diameter classes (channels).
The lower limit of a diameter class is defined by the following relation \cite{Montopolietal2008},
\begin{equation}\label{lowerborder}
d_{c_{i}} = \left[\frac{10^{(1-\alpha(127-i))}}{\gamma}\right]^{\beta},
\end{equation}
where $i$ is the diameter class index, $d_{c_{i}}$ the lower
boundary of the $i$-th class, \mbox{$\alpha$$=$0.014253},
\mbox{$\beta$$=$0.6803}, and \mbox{$\gamma$$=$0.94}. Thus, the
range of observed diameters is $[$0.3mm,5mm$]$.
Eq.~(\ref{lowerborder}) makes the diameter classes uniformly
spaced in a logarithmic scale so that small drops are classified
through finer classes than those used for large drops. 
The interested reader can find a detailed discussion on impact disdrometers and their limitations 
in \cite{JW67,SBE90,SJ94,MFL93,SL95},while more detailed information about the instrument at Chilbolton 
can be found in \cite{Montopolietal2008}.


\end{document}